# An Approach to Handle Big Data Warehouse Evolution

Darja Solodovnikova and Laila Niedrite

Faculty of Computing, University of Latvia, 19 Raina blvd., Riga, Latvia
{darja.solodovnikova, laila.niedrite}@lu.lv

**Abstract.** One of the purposes of Big Data systems is to support analysis of data gathered from heterogeneous data sources. Since data warehouses have been used for several decades to achieve the same goal, they could be leveraged also to provide analysis of data stored in Big Data systems. The problem of adapting data warehouse data and schemata to changes in these requirements as well as data sources has been studied by many researchers worldwide. However, innovative methods must be developed also to support evolution of data warehouses that are used to analyze data stored in Big Data systems. In this paper, we propose a data warehouse architecture that allows to perform different kinds of analytical tasks, including OLAP-like analysis, on big data loaded from multiple heterogeneous data sources with different latency and is capable of processing changes in data sources as well as evolving analysis requirements. The operation of the architecture is highly based on the metadata that are outlined in the paper.

## 1 Introduction

Data warehouses and OLAP methods have been used for several decades to support analysis of structured data sets and, therefore, many solutions to known research problems have been developed in the context of traditional relational database environments. One of such problems is a data warehouse evolution that occurs due to changes in business requirements or data sources or improvements of a data warehouse design.

Recently, due to the increase in the volume and heterogeneity of data that need to be processed and analyzed, Big Data technologies [1] have emerged that use distributed data storage methods and process data in parallel. The demand to analyze data stored in such systems is increasing and one of the analysis options is to use data warehousing. Open Big Data challenges and research directions have been outlined in several recent articles. For example, in [2] the authors state that detecting data and structural changes and handling them in a Big Data system is a complex task that requires further research.

To solve the Big Data evolution problem, we propose an architecture that allows to store and process structured and unstructured data at different levels of detail, analyze them using OLAP capabilities and semi-automatically manage changes in requirements and data expansion. The operation of the architecture components responsible for OLAP analysis and evolution handling is based on the metadata described in the paper.

## 2  Big Data Warehouse Architecture

The idea of the Big Data warehouse architecture was inspired by our previous work on traditional data warehouse evolution framework [3]. We extended the data warehouse evolution framework to the context of Big Data.

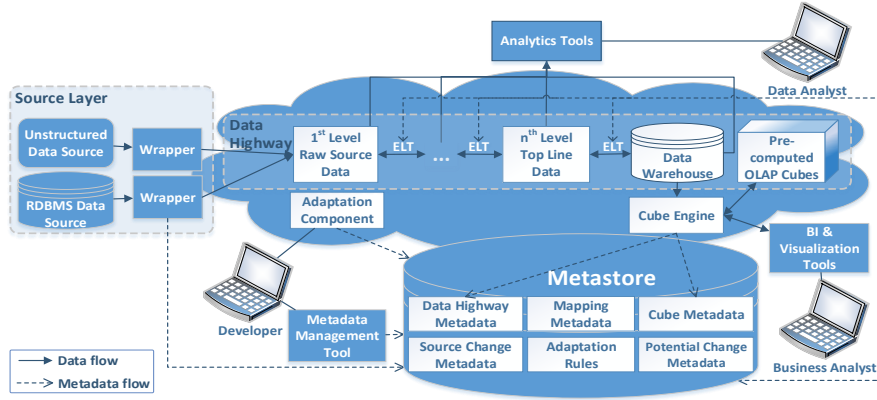

**Fig. 1.** Big Data warehouse architecture

### 2.1  Architecture Components

The proposed architecture consists of several components (Figure 1). In the source layer, wrappers obtain structured and unstructured data from data sources and load them into the system at different rates in their original format. We adopt the idea proposed by [4] to build a highway of data at different levels of latency. Starting from the raw source data, each next level data are obtained from the previous level data and are updated less often. Besides, the latter level data from multiple heterogeneous sources are integrated and aggregated and finally are transformed into a structured data warehouse schema. Since data in the proposed architecture are firstly copied in their original format and transformed at the later stage, ELT (Extract, Load, Transform) processes are responsible for data pre-processing. To gain structured data from unstructured sources, advanced methods such as data mining or sentiment analysis must be performed.

The adaptation component is responsible for handling changes in data sources. The main idea is to generate several potential changes in a data warehouse or other levels of the data highway for each change in a data source and to allow a developer to choose the most appropriate change that must be implemented. To implement certain kinds of changes, the developer needs to supply via the adaptation component additional data that cannot be identified automatically.

We plan to support different kinds of analysis. OLAP cubes may be explored by business analysts in the form of dashboards, charts or other reports and by performing OLAP operations using business intelligence and visualization tools. Since the volume of data stored in the data warehouse may be too large to provide a reasonable performance of data analysis queries, the cube engine component pre-computes various di-mensional combinations and aggregated measures for them. Apart from OLAP operations, data analysts may apply advanced analysis techniques (for example, data mining) by means of utilizing existing analytics tools or implementing ad-hoc procedures.

### 2.2 Metadata Management

One of the most essential components of the proposed architecture is the metastore that incorporates six types of interconnected metadata necessary for the operation of other components of the architecture. Data highway metadata describe semantics and schemata of Big Data stored in the different levels of the system. Cube metadata describe schemata of pre-computed cubes and are leveraged not only during the cube computation process but also for execution of queries. Mapping metadata define the logics of ELT processes. They store the correspondences between data obtained from the sources and data items of the data highway. Information about changes in data sources is accumulated in the source change metadata. Such information may be obtained from wrappers or during the execution of ELT processes. Adaptation rules specify adaptation options that must be implemented for different types of changes. Finally, potential change metadata accumulate proposed changes in the data warehouse schema.

To maintain the information in the metastore, a developer utilizes the metadata management tool. In addition, the metadata management tool allows the developer to initiate changes in the data highway and ELT procedures to handle new or changed requirements for data. The history of chosen changes that are implemented to propagate evolution of data sources, as well as changes performed directly via the metadata management tool, are also maintained in the potential change metadata.

## 3 Related Work

Although, different approaches have been proposed to handle data warehouse evolution in relational database environments, e.g. [5], [6], they cannot be utilized directly to perform adaptation of Big Data systems. Only few Big Data solutions are considering the evolution aspect of Big Data, e.g. [7], however, the objective of the presented system is not data analysis and it does not employ a data warehouse.

Although, there are several studies devoted to the multidimensional analysis of Big Data, e.g. [8], they do not address the problem of Big Data evolution. We found only few researches that consider that problem. In the paper [9] authors discuss a model enrichment process and propose iterative execution of the model design step of the methodology for a construction of a Big Data analysis system. The proposed approach is complimentary to ours. A data warehouse solution for Big Data analysis described in the paper [10] supports two kinds of changes: slowly changing dimensions and fact table schema versions in metadata. Unlike our proposal, the system does not process changes in Big Data sources. An architecture that exploits Big Data technologies and is used for OLAP analytics at LinkedIn is presented in the paper [11]. Authors mention the problem of cube evolution when a new dimension is added to the cube. They handle such changes by means of manual cube schema redefinition and data re-calculation.

# Conclusions and Future Work

In this paper we proposed the data warehouse architecture for supporting Big Data analysis. The unique feature of our proposed architecture is that it is capable of automatically or semi-automatically adapting to changes in requirements or data expansion.

The architecture has not been fully implemented, therefore, our future research directions include a construction of metadata models to describe schemata of the data highway, requirements for data, source data and changes and development of algorithms for automatic and semi-automatic change detection and treatment. For the implementation of the big data warehouse architecture, we intend to utilize the existing tools and technologies as well as to implement the original solutions.

**Acknowledgments.** This work has been partly supported by the ERDF project No. 1.1.1.2./VIAA/1/16/057 and by University of Latvia project No. AAP2016/B032.

# References

1. Thusoo, A., Sarma, J.S.., Jain, N., Shao, Z., Chakka, P., Zhang, N., Antony, S., Liu, H., Murthy, R.: Hive - a petabyte scale data warehouse using Hadoop. In: International Conference on Data Engineering, pp. 996–1005. (2010)
2. Ceravolo, P., Azzini, A., Angelini, M., Catarci, T., Cudré-Mauroux, P., Damiani, E., Mazak, A., Van Keulen, M., Jarrar, M., Santucci, G., Sattler, K., Scannapieco, M., Wimmer, M., Wrembel, R., Zaraket, F.: Big Data Semantics. J Data Semantics 7(2), 65-85 (2018)
3. Solodovnikova, D.: Data Warehouse Evolution Framework. In: Spring Young Researchers Colloquium on Database and Information Systems SYRCoDIS, pp. 4. (2007)
4. Kimball, R., Ross, M.: The Data Warehouse Toolkit: The Definitive Guide to Dimensional Modeling. 3rd edition. John Wiley & Sons, Inc. (2013)
5. Golfarelli, M., Lechtenbörger, J., Rizzi, S., Vossen, G.: Schema versioning in data warehouses: Enabling cross-version querying via schema augmentation. Data & Knowledge Engineering, 59(2) 435-459 (2006)
6. Ahmed, W., Zimányi, E., Wrembel, R.: A Logical Model for Multiversion Data Warehouses. In: Intl. Conf. on Data Warehousing and Knowledge Discovery, pp. 23-34. (2014)
7. Nadal, S., Romero, O., Abelló, A., Vassiliadis, P., Vansummeren, S.: An integration-oriented ontology to govern evolution in big data ecosystems. In: Workshops of the EDBT/ICDT 2017 Joint Conference. (2017)
8. Song, J., Guo, C., Wang, Z., Zhang, Y., Yu, G., Pierson, J.: HaoLap: A Hadoop based OLAP system for big data. Journal of Systems and Software, 102, 167-181. (2015)
9. Tardio, R., Mate, A., Trujillo, J.: An Iterative Methodology for Big Data Management, Analysis and Visualization. In: International Conference on Big Data, pp. 545-550. (2015)
10. Chen, S.: Cheetah: A High Performance, Custom Data Warehouse on Top of MapReduce. VLDB Endowment, 3(2), 1459-1468. (2010)
11. Wu, L., Sumbaly, R., Riccomini, C., Koo, G., Kim, H.J., Kreps, J., Shah, S.: Avatara: OLAP for Web-scale Analytics Products. VLDB Endowment, 5(12), 1874-1877. (2012)